\documentclass[prX,showpacs,superscriptaddress, twocolumn ]{revtex4}

\usepackage{graphicx}
\usepackage{amsmath}
\usepackage{amsfonts}
\usepackage{amssymb}
\usepackage{epsf}
\usepackage{hyperref}
\usepackage{pstricks,pst-coil,pst-fill,pst-plot}

\begin{document}

\title{Band gap structures for matter waves}

\author{F. Damon}
\affiliation{Laboratoire de Physique Th\'eorique (IRSAMC), Universit\'e de Toulouse 
(UPS), 31062 Toulouse, France} 
\affiliation{CNRS, LPT UMR 5152 (IRSAMC), 31062 Toulouse, France} 

\author{G. Condon} 

\affiliation{Universit\'e de Toulouse ; UPS ; Laboratoire Collisions Agr\'egats R\'eactivit\'e, IRSAMC ; F-31062 Toulouse, France} 
\affiliation{CNRS ; UMR 5589 ; F-31062 Toulouse, France}

\author{P. Cheiney}

\affiliation{Universit\'e de Toulouse ; UPS ; Laboratoire Collisions Agr\'egats R\'eactivit\'e, IRSAMC ; F-31062 Toulouse, France} 
\affiliation{CNRS ; UMR 5589 ; F-31062 Toulouse, France}
\affiliation{ICFO-Institut de Ciencies Fotoniques, Mediterranean Technology Park, 08860 Castelldefels (Barcelona), Spain}

\author{A. Fortun}

\affiliation{Universit\'e de Toulouse ; UPS ; Laboratoire Collisions Agr\'egats R\'eactivit\'e, IRSAMC ; F-31062 Toulouse, France} 
\affiliation{CNRS ; UMR 5589 ; F-31062 Toulouse, France}

\author{B. Georgeot}

\affiliation{Laboratoire de Physique Th\'eorique (IRSAMC), Universit\'e de Toulouse 
(UPS), 31062 Toulouse, France} 
\affiliation{CNRS, LPT UMR5152 (IRSAMC), 31062 Toulouse, France} 

\author{J. Billy}

\affiliation{Universit\'e de Toulouse ; UPS ; Laboratoire Collisions Agr\'egats R\'eactivit\'e, IRSAMC ; F-31062 Toulouse, France} 
\affiliation{CNRS ; UMR 5589 ; F-31062 Toulouse, France}

\author{D. Gu\'ery-Odelin}

\affiliation{Universit\'e de Toulouse ; UPS ; Laboratoire Collisions Agr\'egats R\'eactivit\'e, IRSAMC ; F-31062 Toulouse, France} 
\affiliation{CNRS ; UMR 5589 ; F-31062 Toulouse, France}

\date{\today}

\begin{abstract}
Spatial gaps correspond to the projection in position space of the gaps of a periodic structure whose envelope varies spatially. They can be easily generated in cold atomic physics using finite-size optical lattice, and provide a new kind of tunnel barriers which can be used as a versatile tool for quantum devices. 
We present in detail different theoretical methods to quantitatively describe these systems, and show how they can be used to realize in one dimension matter wave Fabry-Perot cavities. We also provide experimental and numerical results that demonstrate the interest of spatial gaps structures for phase space engineering. We then generalize the concept of spatial gaps in two dimensions and show that this enables to design multiply connected cavities which generate a quantum dot structure for atoms or allow to construct curved wave guides for matter waves. 
At last, we demonstrate that modulating in time the amplitude of the periodic structure offers a wide variety of possible atom manipulations including the control of the scattering of an incoming wave packet, the loading of cavities delimited by spatial gaps, their coupling by multiphonon processes or the realization of a tunable source of atoms. This large range of possibilities offered by space and time engineering of optical lattices demonstrates the flexibility of such band gap structures for matter wave control, quantum simulators and atomtronics.
\end{abstract}


\maketitle


\section{Introduction}

Metamaterials are a fast-developing research field with a wide variety of applications \cite{Metamaterials}. A famous example in optics is provided by photonic band gap structures that exploit multi-wave interferences \cite{Nanophotonics}. Optical components such as wave guides, bending light, micro-resonators or filters can be realized by introducing periodicity defects at suitably selected spots within a crystal. By pairing them it is now possible to design photonic circuits.

Inspired by those developments in optics, a few studies have proposed to investigate their counterpart in atom optics \cite{Santos1,Santos2,Carusotto1,Carusotto2,Sanpera}. The first element that was envisioned was the Bragg or multilayer dielectric mirror in which the layers were provided by a laser light interference pattern i.e. a periodic succession of   dark and bright layers with a typical size of hundreds of nm \cite{Santos1,Santos2}. In contrast with optics, the multiple matter wave interferences are not used here to improve the reflectivity but to gain on velocity selectivity and to design ultra narrow and tunable velocity filters \cite{Carusotto1,Carusotto2}. Indeed, many quantum devices require specific shaping or momentum-selective filter of an atomic wave packet \cite{atomtr1,atomtr1b,atomtr2,atomtr2b,atomtr3,atomtr4,atomtr4b,atomtr5,atomtr6,Damon2014}. 

The experimental realization of such a mirror was performed by studying the scattering of a guided Bose-Einstein condensate on a one-dimensional and finite-size optical lattice \cite{Fabre11}: the band structure was directly probed and a wide variety of velocity filters (notch, band pass, high and low pass) has been demonstrated. The envelope of the optical lattice projects in real space the gaps of the band structure. The corresponding spatial gaps generate a new type of atomic tunnel barriers. Crossing such a barrier in real space amounts to performing a Landau-Zener transition between two adjacent bands \cite{LaL77, Zen32}. In \cite{dgoepl13}, a matter wave cavity delimited by two spatial gaps has been investigated. Among many prospects, the shaping of the envelope using for instance Spatial Light Modulators (see e.g. \cite{slm}) should offer the possibility to realize a mode-locked atom laser. 

In contrast with structured materials used in optics, the optical lattice depth can be modified in time. This possibility was exploited experimentally in \cite{Cheiney13} to realize tunable velocity filters and mirrors. This brings forward a new tool for the emerging field of atomtronics \cite{atomtr1,atomtr1b,atomtr2,atomtr2b,atomtr3,atomtr4,atomtr4b,atomtr5,atomtr6,circuits}: the components that will be combined to realize an atomic circuit could have a tunable functionality.

In this article, we detail different methods to calculate in practice the properties of spatial gaps and propose several possible applications of this concept, from the implementation and observation of complex quantum phenomena to the construction of versatile tools for atomtronics. To this aim, we study in detail different setups in one or two dimensions using static or time-modulated potentials. 
 After a short reminder about the Bloch band formalism (Sec.~\ref{secevanescent}A), the Hill's method is presented in Sec.~\ref{secevanescent}B. This mathematical trick provides the exact imaginary part of a wave vector $\kappa$ inside the gaps of a one-dimensional, infinite and uniform periodic potential. The characterization of those evanescent matter wave modes plays a key role in the understanding of spatial gaps. However, the exact mathematical treatment is restricted to one-dimensional system. In Sec.~\ref{secevanescent}C, we use those exact results to validate a systematic perturbative approach to evanescent modes that can be readily generalized to higher dimensions. Those results about evanescent modes are then used to define precisely the concept of spatial gaps within a locally uniform potential depth approximation and discuss their properties (Sec.~\ref{secspatialgaps}). In Sec.~\ref{sec4}, we discuss the dynamics of a wave packet in the presence of two spatial gaps (Fabry-Perot like device) or more. We also demonstrate experimentally the interest of such setups for phase space engineering.
 In Sec.~\ref{highdim} we address the generalization of the previous ideas to two dimensions leading to the design of simply and multiply connected matter wave cavities and of curved wave guides. Section \ref{floquet} is devoted to the generalization to the time domain of the concept of spatial gaps. We show how multiple cavities can be designed in particular in the wing of the envelope of the optical lattice and how their absolute and relative population can be controlled by multiphonon processes. 
\section{Gaps and evanescent modes in an optical lattice}
\label{secevanescent}

\subsection{Bloch formalism}

Before tackling the detailed treatment of spatial gaps, we first provide hereafter a short reminder on the Bloch formalism for quantum particles evolving in a periodic structure.
Consider a particle of mass $m$ in a one-dimensional periodic potential of period $d$, $U(z+d)=U(z)$.
The stationary Schr\"{o}dinger equation that describes the dynamics in such a potential reads:
\begin{equation}
H\psi(z)=\left(\frac{p^2}{2m}+U(z)\right)\psi(z)=E\psi(z).
\label{Eq:Schrodinger_indep_temps}
\end{equation}
The Bloch theorem states that the eigenstates of a periodic Hamiltonian may be written as the product of a plane wave function with wave-vector $k$ and a function $u_{n,k}(z)$ that has the same periodicity as that of the potential $U$ \cite{Ashcroft}:
\begin{equation}
\psi_{n,k}(z)=e^{ikz}u_{n,k}(z) \;\; \textrm{with} \;\; u_{n,k}(z+d)=u_{n,k}(z).
\end{equation}
Except for $k=0$, these Bloch states are propagating states.
The eigenenergies associated with $\psi_{n,k}$, $E_n(k)$, are periodic with period $k_{\rm R}=2\pi/d$: $E_n(k)=E_n(k+k_{\rm R})$. By inserting $\psi_{n,k}$ in Eq.~(\ref{Eq:Schrodinger_indep_temps}), we deduce that the functions $u_{n,k}$ are the eigenstates of an effective $k$-dependent Hamiltonian:
\begin{equation}
H_k = \frac{(p+\hbar k)^2}{2m}+U(z).
\end{equation}
Using this eigenvalue equation and the boundary condition $u_{n,k}(z)=u_{n,k}(z+d)$, the signification of the band index becomes clear. $E_n(k)$ are the eigenvalues of an effective Hamiltonian in a box with periodic boundary conditions. As such, we expect a quantification of the energies that correspond to the discrete band index $n$.
Because of the periodicity of the Bloch functions $u_{n,k}$, we can limit ourselves to the first Brillouin zone, i.e. $k \in [-k_{\rm{R}}/2,k_{\rm{R}}/2]$. By expanding the periodic Bloch functions $u_{n,k}$ as Fourier series, we get
\begin{equation}
\psi_{n,k}(z)=e^{ikz} u_{n,k}(z)=\sum_\ell v_{k+\ell k_{\rm{R}}} e^{i(k+\ell k_{\rm{R}})z}.
\label{Eq:state_fourier}
\end{equation}
For a given $k$, working out the Bloch state $\psi_{n,k}(z)$ amounts to finding the coefficients $v_{k+\ell k_{\rm{R}}}$. Similarly, the potential $U$ can be Fourier expanded:
\begin{equation}
U(z)=\sum_p \tilde U_p e^{ipk_{\rm{R}} z}.
\label{Eq:potential_fourier}
\end{equation}
Combining Eqs.~(\ref{Eq:state_fourier}) and (\ref{Eq:potential_fourier}) with the Schr\"{o}dinger equation (\ref{Eq:Schrodinger_indep_temps}), we infer the following set of coupled equations:
\begin{equation}
\frac{\hbar^2}{2 m}\left(k+\ell k_{\rm{R}}\right)^2v_{k+\ell k_{\rm{R}}} +\sum_p \tilde U_p v_{k+(\ell-p) k_{\rm{R}}}=E v_{k+\ell k_{\rm{R}}}.
\label{Eq:matrix}
\end{equation}
To find the eigenvalues $E_n(k)$, we truncate this infinite linear system and solve the resulting finite size matrix equation. Consider an attractive lattice of potential 
\begin{equation}
U(z)=-U_0 (\cos(k_{\rm{R}} z)+1)/2,
\label{potattr}
\end{equation} 
where the $U_0>0$ is the depth of the lattice.
The matrix ${\mathbf M}$ that provides the eigenvalues through ${\mathbf M}\cdot {\mathbf V}=E{\mathbf V}$ has a simple band structure with
\begin{equation}
{\mathbf M}=\begin{pmatrix}
 b_{-N}& u & & &   \\
 u & b_{-N+1} &  u &  & \\
 &\ddots &\ddots & \ddots&  &\\
  & & u&  b_{N-1}& u  \\
  & & & u & b_N
\end{pmatrix},\label{matrix2}
\end{equation}
where $u=-U_0/4$, $b_\ell=E_{\rm{R}}\left(k/k_{\rm{R}}+\ell\right)^2-U_0/2$ and $E_{\rm{R}}=\hbar^2k_{\rm R}^2/2m$. In this way, one finds for each $k$ in the Brillouin zone, a discrete set of energies $E_n(k)$. By plotting these eigenvalues for all values of $k$, we obtain the energy bands that are separated by energy gaps. The border of the gaps are obtained either at $k=0$ or $k=\pm k_{\rm R}/2$. In the energy gap there exist solutions of the eigen-equation for complex values of the wave vector $k$. They are associated with evanescent modes and are usually invoked in solid state physics only when dealing with surfaces and junctions. They are at the heart of this article. In the next subsection, we provide a reminder on Hill's method to calculate the expression of those complex wave vectors. We then propose an alternative method to calculate them approximatively through a perturbative approach. This latter method has the advantage of being more physical and can be directly generalized to higher dimensions.

\subsection{Hill's method}\label{Hill}

Using the potential (\ref{potattr}), the Schr\"odinger equation is nothing but a Mathieu equation. Calculating the complex wave vector associated with an energy lying in a band gap amounts to calculate the so-called Mathieu characteristic exponent $\kappa\in\mathbb{C}$. In order to determine them, we detail hereafter the Hill method following \cite{whittaker_watson,Strang,McLachlan_mathieu}. 

The eigenvalue matrix equation can be recast in the form ${\mathbf A}(\kappa;E,U_0)\cdot{\mathbf V}=0$ with
\begin{equation}
{\mathbf A}(k;E,U_0)=\begin{pmatrix}
 1& \xi_{-N} & & &   \\
 \xi_{-N+1} & 1 &  \xi_{-N+1} &  & \\
 &\ddots &\ddots & \ddots&  &\\
  & & \xi_{N-1}&  1& \xi_{N-1}  \\
  & & & \xi_{N} & 1
\end{pmatrix}
\label{determinant}
\end{equation}
where $\xi_\ell=(-U_0/4E_{\rm{R}})/((\kappa/k_{\rm{R}}+\ell)^2-\zeta)$ and $\zeta=(U_0/2+E)/E_{\rm{R}}$. Nontrivial solutions are obtained for $\Delta(\kappa)=\det(A)=0$. In the limit $N \rightarrow \infty$, $\Delta(\kappa)$ is periodic with period $k_{\rm{R}}$. We can thus restrict the domain of Re[$\kappa$] to the domain: $0\le \text{Re}[\kappa]\le k_{\rm{R}}$. Since each $\xi_\ell(\kappa)$ appears only once per line, $\Delta$ is a sum of products of the $\xi_\ell(\kappa)$. The $\xi_\ell$ are analytical functions of $\kappa$ except at poles which are of finite order (meromorphic functions); $\Delta$ is thus analytical except at $\kappa/k_{\rm{R}}=\sqrt{\zeta}-\ell$. Since each function $\xi_\ell$ appears only once in $\Delta$, all the poles of $\Delta(\kappa)$ are simple and located at $\kappa/k_{\rm{R}}=\sqrt{\zeta}-\ell$. 
 
The trick of Hill's solution consists in considering the function \cite{whittaker_watson}
\begin{equation}
D(\kappa)=\frac{1}{\cos(2 \pi \kappa/k_{\rm{R}})-\cos(2 \pi \sqrt{\zeta})}
\end{equation}
that has exactly the same poles as $\Delta$. One can thus choose a function $C(\kappa)$ such that the function defined by 
\begin{equation}
\Theta(\kappa)=\Delta(\kappa)-C(\kappa)\cdot D(\kappa)
\end{equation}
has no singularities. In the interval $[0,k_{\rm{R}}]$, $\Delta$ has only one pole and the function $C$ can be chosen constant and equal to the ratio between the residues of the functions $\Delta$ and $D$ at this unique pole. With such a choice for $C$, the function $\Theta$ is analytical on the whole complex plane (holomorphic function). It must then be a constant according to Liouville's theorem.

To determine the value of $C$, we use the limit $\kappa\rightarrow +i\infty$. We find $D(\kappa\to +i\infty)=0$ and $\Delta(\kappa\to +i\infty)=1$, and deduce that $\Theta(\kappa\rightarrow +i\infty)=1$. As a result, we obtain the value of $C=(\Delta(\kappa)-1)/D(\kappa)$.
Using $D(0)=1/(1-\cos(2\pi\sqrt{\zeta}))$, we infer its explicit expression
\begin{equation}
C=(\Delta(0)-1)\cdot(1-\cos(2\pi\sqrt{\zeta})).
\end{equation}
From $\Delta(\kappa)=0$ (non-trivial solution), we finally get the equation fulfilled by $\kappa$ for a given energy $E$ (through the parameter $\zeta$):
\begin{equation}
\cos(2\pi \kappa/k_{\rm{R}})=1-\Delta(0)(1-\cos(2 \pi \sqrt{\zeta})).
\label{eqnl}
\end{equation}
Real solutions of this equation provide the equations for the energy bands $E(k)$, while those with an imaginary part account for the evanescent modes in the gaps. 

\subsection{The perturbative approach} \label{section:perturbative}

The previous approach is exact but requires to solve numerically the nonlinear equation (\ref{eqnl}). It is possible to work out a perturbative treatment of the evanescent modes, for two important purposes: to get an explicit formula for the imaginary wave vector valid in the low lattice depth limit and to use it as a starting point for obtaining the imaginary wave vectors for 2D or 3D optical lattices.

To investigate perturbatively the first gap, we use the two mode approximation for the potential (\ref{potattr}). It consists in reducing the infinite set of equations (\ref{Eq:matrix}) to two coupled equations involving the modes at the edge of the Brillouin zone $k=\pm k_{\rm R}/2$ \cite{Smith}:
\begin{eqnarray}
&(E_{\rm R}/4 - U_0/2)v_{k_{\rm R}/2} -(U_0/4)v_{-k_{\rm R}/2}=E v_{k_{\rm R}/2}, \nonumber \\
&-(U_0/4)v_{k_{\rm R}/2} + (E_{\rm R}/4 - U_0/2)v_{-k_{\rm R}/2}=E v_{-k_{\rm R}/2}. \label{bragg}
\end{eqnarray}

\begin{figure}[h!]
\centering
\includegraphics[width=8cm]{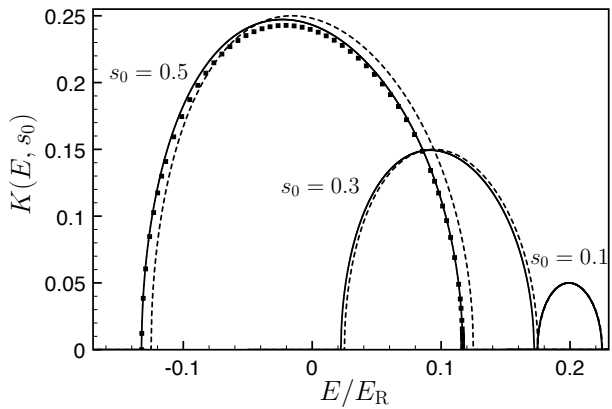}
\caption{Imaginary part of the wave vector normalized to $k_{\rm R}/2$, $K(E,s_0)$, in the first band gap for $s_0=0.1$, $s_0=0.3$ and $s_0=0.5$: exact calculation (solid line), perturbative calculation to the lowest order (dashed line) and perturbative calculation pushed to the next order (black square).}
\label{fig1}
\end{figure}

Introducing the reduced depth parameter $s_0=U_0/E_{\rm R}$, the gap borders deduced from the previous set of equations are given by $E_\pm=(E_{\rm R}/4)(1-2s_0\pm s_0)$. To determine the imaginary part of wave vectors inside the corresponding band gap, we set the energy equal to $E_x=(E_{\rm R}/4)(1-2s_0 +  s_0x)$ with $-1<x<1$. We then search for a two-mode solution with $\kappa_\pm=\pm (1\pm iK)(k_{\rm R}/2)$ where $K$ depends on $x$ and $s_0$:
\begin{eqnarray} 
&& (E_{\rm R}(1+iK)^2 -E_{\rm R} - xU_0)v_{k_{\rm{R}/2}} - U_0v_{-k_{\rm{R}/2}} = 0, \nonumber \\  
&& -U_0v_{k_{\rm{R}}/2} + (E_{\rm R}(1-iK)^2 -E_{\rm R} - xU_0)v_{-k_{\rm{R}}/2}  = 0. 
\end{eqnarray}
This set of equations has a solution if $K$ obeys the equation
\begin{equation}
K^4 + K^2(4+2xs_0) + s_0^2(x^2-1)=0
\end{equation}
whose solution is given by
\begin{eqnarray} 
K(x,s_0) & =&  \sqrt{-2-s_0x+\sqrt{4+s_0^2+4s_0x}}\nonumber \\  
& \simeq & s_0\sqrt{1-x^2}/2,\text{ for }|s_0|\ll1.\label{imkappa}
\end{eqnarray}
Figure~\ref{fig1} summarizes the comparison between the exact method and the perturbative approach detailed above for different lattice depths $s_0=0.1$, $s_0=0.3$ and $s_0=0.5$. We observe that the maximum of the bell shaped curve for $K$ is indeed on the order of $s_0/2$ as expected from the perturbative treatment. As expected the accuracy of the perturbative method based on the two mode approximation decreases as $s_0$ increases. In appendix A, we discuss more systematically the validity of the two mode approximation.

The strategy to get a higher accuracy is quite clear, we have to add extra coupled modes. In the perturbative approach presented in the previous paragraph we have only taken into account the coupling between two modes $v_{-k_{\rm{R}}/2}$ and $v_{k_{\rm{R}}/2}$. To get a better approximation we can repeat the previous argument with four modes: $v_{-3k_{\rm{R}}/2}$, $v_{-k_{\rm{R}}/2}$, $v_{k_{\rm{R}}/2}$ and $v_{3k_{\rm{R}}/2}$. The result is shown in Fig.~\ref{fig1} for $s_0=0.5$. We already obtain a quite good agreement with the exact result (relative distance below 1.7 \%). Increasing further the number of coupled modes will increase the convergence towards the exact result. As expected for such a perturbative approach, to reach a given accuracy the number of modes that have to be taken into account increases with the potential depth ($s_0$ parameter). We have also extended this approach to the second gap in Appendix B.

\section{Spatial gaps as tunnel barriers for matter waves}
\label{secspatialgaps}

\begin{figure}[h!]
\centering
\includegraphics[width=8cm]{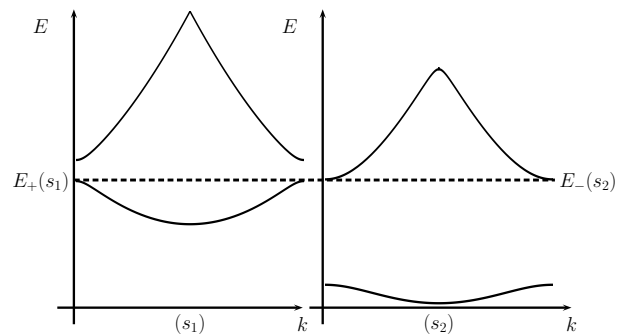}
\caption{Bloch band distributions in the first Brillouin zone for two different potential depths $s_1$ and $s_2>s_1$. The pseudo-energy (dashed line) is fixed so that $E_+(s_1)=E_-(s_2)$.}
\label{figs1s2}
\end{figure}

The spatial variation of the envelope of an optical lattice projects the band gaps in real space. The latter are therefore referred to as spatial gaps. In the following we consider the first gap of a lattice for which the scale of variation of the envelope is large compared to the lattice period. Under this assumption, one can consider  the lattice locally uniform at each position. For a given pseudo-energy $E$ and potential depth $s$, we consider the closest gap at an energy below $E$. There are two values of the depth $s_1<s$ and $s_2<s$ such that $E_+(s_1)=E_-(s_2)=E$ with $s_1=s(z_1)$ and $s_2=s(z_2)$ (see Fig.~\ref{figs1s2}). In the interval $[z_1;z_2]$, the Mathieu exponent $\kappa$ acquires an imaginary part. The size of the corresponding spatial barrier is $\delta z=|z_1-z_2|$. 
This virtual barrier thus acts as a regular tunnel barrier in the energy range spanned by the gap. When crossing such a barrier, the atom undergoes a Landau-Zener transition between two adjacent bands \cite{LaL77, Zen32}.

To calculate the transmission probability $T(E)$, through the barrier, we shall calculate the value of the imaginary part Im$[\kappa(z;E)]$ in the interval $[z_1;z_2]$:
\begin{equation}
T(E)=\exp \left(  -2\int_{z_1}^{z_2} \text{Im}[\kappa(z;E)]{\rm d}z \right),
\label{tdeE}
\end{equation}
 In this interval, the value of the depth $s$ varies with space but the pseudo-energy, $E$, remains fixed. For $s$ in the interval $[s_1;s_2]$, the value of $x$ is set by the energy:
\begin{equation}
E= \frac{E_-(s)+E_+(s)}{2} + x\frac{E_+(s)-E_-(s)}{2}. 
\end{equation}
where $x$ depends on $z$ through the local depth parameter $s(z)$.
At the scale of the width of the barrier, we can assume that the envelop varies linearly:
$s(z)=s(z_1) + (z-z_1)\left( \partial_z s  \right)_{z_1}$. We deduce

\begin{equation}
T(E)=\exp \left(  - \frac{f(E)}{\left( \partial_z s  \right)_{z_1}}\right)
\end{equation}
where $f(E)$ is a function that depends only on the energy:

\begin{equation}
f(E)=2\int_{s_1}^{s_2}\text{Im} [\kappa(E,s)]{\rm d}s .
\end{equation}

This exact form therefore exhibits a probability of transmission that decreases exponentially with the inverse gradient of the envelope. The smaller the gradient, the larger the tunnel barrier width. This gradient therefore appears as a parameter that can be tuned experimentally to set the transmission tunnel probability.

This general formalism can be applied in combination with the perturbative approach of Sec.~\ref{secevanescent}C to get explicit formulas for the transmission probability.  In the perturbative limit, we have $\text{Im}[\kappa(e,s)]=k_{\rm R}\sqrt{s^2-(4e-1+2s)^2}/4$ (using Eq.(\ref{imkappa})), $s_1=(1-4e)/3$ and $s_2=3s_1$ (where $e=E/E_{\rm R}$), and find for $e\le1/4$.
\begin{equation}
f(e)=k_{\rm R}\frac{\pi}{24\sqrt3}(1-4e)^2.
\end{equation}
It is worth noticing that the energy dependence of this transmission probability is quite different from that obtained for a real repulsive potential barrier. This turns out to be an advantage of this kind of tunnel barrier since one can achieve pretty high transmission probability for a single barrier as experimentally demonstrated in \cite{dgoepl13}. This is a feature that is difficult, if not impossible, to obtain using an optimally focussed blue detuned laser to realize a repulsive barrier for atoms because of the diffraction limit \cite{Billy8}.

\section{Spatial gap structures in one dimension}
\label{sec4}
The preceding section considered a spatial gap as an isolated barrier. However, it is quite easy to engineer systems where several spatial gaps are present at different locations. 
In this section, we address the new features that emerge when one consider a pair of spatial gaps and discuss the case of multiple barriers.
We also show that this type of device enables to engineer the phase space distribution of wave packets and demonstrate it experimentally.
 
\subsection{Matter wave Fabry-Perot cavity}
\label{secfpcavity}
\begin{figure}[h!]
\centering
\includegraphics[width=8cm]{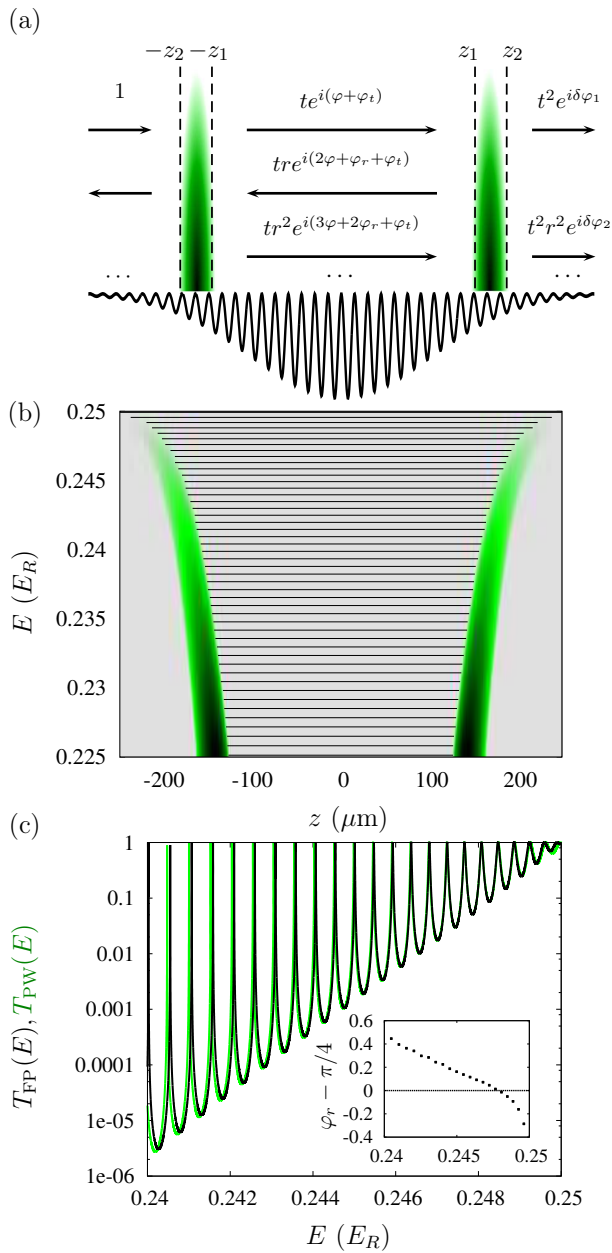}
\caption{(Color online) (a) Optical lattice with a Gaussian envelope showing two spatial gap barriers in the spaces $[-z_2,-z_1]$ and $[z_1,z_2]$. (b) Density plot of Im($k$) associated to the potential (\ref{potential}), as a function of the position and the energy. The solid black lines correspond to the values of the resonances of this cavity  (see text).\label{diagram} (c) Black line: Exact solution for the transmission, $T_{\rm PW}$, as a function of the energy $E$. Green/grey line: Transmission $T_{\rm FP}$ calculated by the semiclassical approach (\ref{transFP}) including a global offset ($\varphi_r=\pi/4$). Inset: Phase difference $(\varphi_r-\pi/4)$ between the semiclassical model and the exact one as a function of energy.\label{transmission}}
\end{figure}

Consider a periodic potential with a Gaussian envelope as used experimentally in \cite{Fabre11}  (Fig.~\ref{diagram}(a)):
\begin{equation}
U(z)=-V_0(z)\sin^2\left(\frac{\pi z}{d}\right)\label{potential}
\end{equation}
with $V_0(z)=U_0\exp\left(-2z^2/w^2\right)$.
This Gaussian shape results from the envelopes of the two Gaussian laser beams that generate the optical lattice. The parameter $U_0>0$ is proportional to the intensity of the laser beams and the lattice spacing $d$ is proportional to the laser wavelength. Hereafter, we choose the typical experimental values $w=140\ \mu\text m$, $d=0.65\ \mu$m  and a potential depth $U_0=E_{\rm R}/2$. 

This potential is symmetric and therefore produces pairs of symmetric spatial gaps. In this section, we focus our study on a single pair of spatial gaps. Each spatial gap acts as a matter wave mirror of energy-dependent reflectivity (see Fig.~\ref{diagram}(b)). As such, the potential experienced by atoms is reminiscent of that of a Fabry-Perot cavity \cite{dgoepl13}. 

The complex amplitude, $\cal A$, of the output wave is obtained by adding the contribution of the multiple reflections of matter waves inside the cavity and reads (see Fig.~\ref{transmission}(a))
\begin{equation}
{\cal A} = t^2e^{i\delta \varphi_1} + t^2r^2e^{i\delta \varphi_2} +  t^2r^4e^{i\delta \varphi_3} +...
\end{equation}
with $\delta \varphi_1 = 2\varphi_t+\varphi$, $\delta \varphi_2 = 2\varphi_t + 3\varphi + 2 \varphi_r$, ... For sake of clarity, the letter $t$ (resp.~$r$) refers to the modulus of the transmission (reflection) amplitude, the corresponding phase is captured in the term $\varphi_t$ (resp. $\varphi_r$) \cite{Note}.
To determine the transmittance function, $T_{\rm FP}= |{\cal A}|^2$, of this Fabry-Perot like device, we square the modulus of the amplitude: 
\begin{align}
T_{\rm FP}(E) & =\left|t^2e^{2i\varphi_t}e^{i\varphi(E)}\left(1+\sum_{n=1}^\infty r^{2n}{\rm e}^{2in(\varphi(E)+\varphi_r)}\right)\right|^2\nonumber\\
&= \left[1+4\left(\frac{1}{T^2(E)}-\frac{1}{T(E)}\right)\sin^2(\varphi(E)+\varphi_r)\right]^{-1}.\label{transFP}
\end{align}
The transmission probability $T_{\rm FP}(E)$ is related to the transmission coefficient by $T(E)=t^2$ (an explicit expression is given in Eq.~(\ref{tdeE})). The phase accumulated over the cavity length can be calculated by a semiclassical approach:
\begin{equation}
\varphi(E)=\frac{1}{2}\oint k\ {\rm d}z=\int_{-z_1}^{z_1}\text{Re}[\kappa(z;E)]\ {\rm d}z.\label{phi}
\end{equation}
We consider a single path from the first barrier to the second, this is the reason why a factor $1/2$ appears in the previous formula.
As expected, the phase of the transmission amplitude, $\varphi_t$, is washed out in the transmission probability.  
However, the phase of the reflection wave, $\varphi_r$, is responsible for a correction to the semiclassical formula that slightly depends on the energy (see below).

 In Fig.~\ref{transmission}(c), we compare the transmission $T_{\rm FP}(E)$ obtained from our semiclassical approach with the transmission $T_{\rm PW}(E)$ obtained by solving numerically the stationary Schr\"odinger equation using plane waves (referred to as the exact result in the following). To obtain an almost perfect agreement for the amplitude and without loss of generality, we have included a global $\pi/4$ offset in the phase of the semiclassical model. A more careful analysis reveals that a small energy drift of the position of the transmission peaks remains. The inset of Fig.~\ref{transmission}(c) represents the corresponding energy-dependent phase difference, $\varphi_r-\pi/4$, between the semiclassical model and the exact solution of the Schr\"odinger equation (the shift correction is on the order of 1\% (0.03 rad per peak)). This small residual shift confirms the good predictability of the semiclassical approximation.

\begin{figure}[h!]
\centering
\includegraphics[width=8cm]{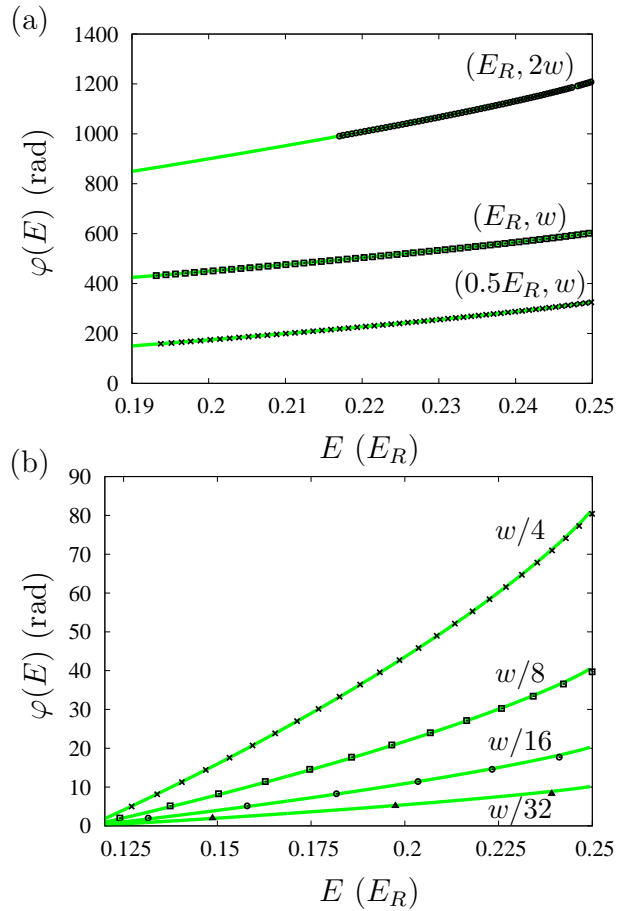}
\caption{(Color online) Solid green/grey line: numerical integration of the semiclassical phase (\ref{phi}). Black dots: position of the resonances deduced from the exact solution shown in Fig.~\ref{transmission}(c). We compare those two quantities for different potential depths (a) and widths (b). \label{phase}}
\end{figure} 
 
To perform a direct comparison of the energy dependence of the phase acquired in transmission through the cavity, we proceed in the following manner: (i) we identify precisely each transmission peak of the Schr\"odinger simulation and (ii) we assign a $\pi$ phase shift between two successive peaks. In Fig.~\ref{phase}, we plot the corresponding cumulated phase (black dots) and compare it with the semiclassical evaluation of the phase $\varphi(E)$, performed through Eq.~(\ref{phi}) using the Hill's method (see subsection \ref{Hill}). We find once again a very good agreement between the two methods \cite{Note2}. As intuitively expected, the phase strongly depends on the waist $w$ of the gaussian envelope (Fig.~\ref{phase}(b)). It is therefore possible to tune the number of resonances by changing the cavity width, in accordance with what could be expected from standard phase space volume arguments. For example for a waist of 140 $\mu$m/32 $\simeq$ $4$ $\mu$m, only 3 resonances remains. Remarkably, the local envelope approximation at the heart of the semiclassical approach still holds when the potential contains only a few lattice periods.\\

\begin{figure}[h!]
\centering
\includegraphics[width=8cm]{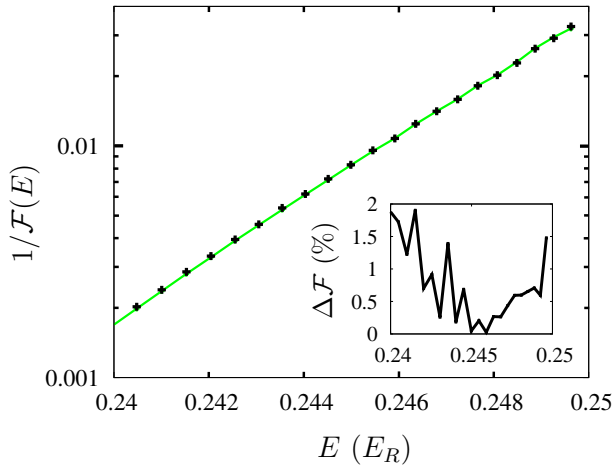}
\caption{(Color online) Inverse of the finesse as a function of the energy, obtained from the exact solution $T_{\rm PW}(E)$ (black dots) and from the semiclassical calculation $T_{\rm FP}(E)$ (green/grey line). Inset: relative error, $\Delta\mathcal{F}(E)=|\mathcal{F}_{\rm PW}-\mathcal{F}_{\rm FP}|/\mathcal{F}_{\rm FP}$. The parameters are the same as in Fig.~\ref{transmission}. \label{finesse}}
\end{figure}

In contrast with the usual optical Fabry-Perot cavity, a cavity based on two spatial gaps has a finesse that strongly depends on the energy:
\begin{equation}
\mathcal{F}(E)=\frac{\Delta_E}{\sigma_E}=\frac{\pi}{2}\left[\arcsin\left(\frac{1}{2\sqrt{1/T^2(E)-1/T(E)}}\right)\right]^{-1},\label{finM}
\end{equation}
where $\Delta_E$ accounts for the energy difference between two successive peaks, and $\sigma_E$ is the standard deviation of each peak. Figure~\ref{finesse} provides a comparison of the finesse calculated from Eq.~(\ref{finM}) with that obtained from the integration of the Schr\"odinger equation. The finesse decreases (almost exponentially) with the energy. The quality of the agreement is quantitatively evaluated by computing the relative error $\Delta\mathcal{F}(E)$ between the two approaches. The inset of Fig.~\ref{finesse} shows that this error remains lower than $2\%$, in the considered energy range and tends to increase slightly for lower energies.

The semiclassical expression for the transmission provides a simple way to evaluate the lifetime of a wave packet trapped in the cavity at a mean energy that coincides with that of a resonance. Indeed, the expansion of the phase about the resonance ($\varphi(E)\simeq\pi$) reads
\begin{equation}
\sin(\varphi(E)+\delta \varphi(E))\approx-\frac{\partial\varphi}{\partial E}\delta E=\frac{\tau\delta E}{\hbar},
\end{equation}
where the semiclassical time $\tau$ corresponds to the time required to travel back and forth in the cavity at that energy
\begin{equation}
\tau=-2\hbar\int_0^{z_1}\frac{\partial\text{Re}[\kappa(z;E)]}{\partial E}\ {\rm d}z=-2\int_0^{z_1}\frac{ {\rm d}z}{v(z;E)}
\end{equation}
with $v(z;E)=(1/\hbar)\partial_kE$ is the semiclassical velocity. The expansion of the transmission about the resonance reads 
\begin{equation}
T_{\rm FP}(E)\simeq\left[1+\frac{4}{1-T(E)}\frac{\tau^2}{\hbar^2}(\delta E)^2\right]^{-1},
\end{equation}
from which we infer the time decay, $\tau_{0}$, 
\begin{equation}
\tau_{0}\simeq \frac{\hbar}{2\delta E}\sqrt{1-T(E)}.
\end{equation}

More generally, the size of the cavity provides an energy scale related to the difference of energy between two successive resonances, $\Delta_E$. The finesse of a given resonance has an energy width equals to $\hbar/\tau_{0}$. For parameters such that the quality factor, $\Delta_E\tau_{0}/\hbar$, is large, this system realizes a true Fabry-Perot cavity for matter waves. Let us finally emphasize that the level spacing and the decay rate can also be engineered for a fixed cavity size by renormalization of the mass using an extra superimposed optical lattice \cite{Carusotto2}. Such cavities are good candidates to observe the atom blockade effect in close analogy with its electronic counterpart (Coulomb blockade)\cite{blockade}.

\subsection{Multiple barrier landscape}
\label{landscape}

\begin{figure}[h!]
\centering
\includegraphics[width=8cm]{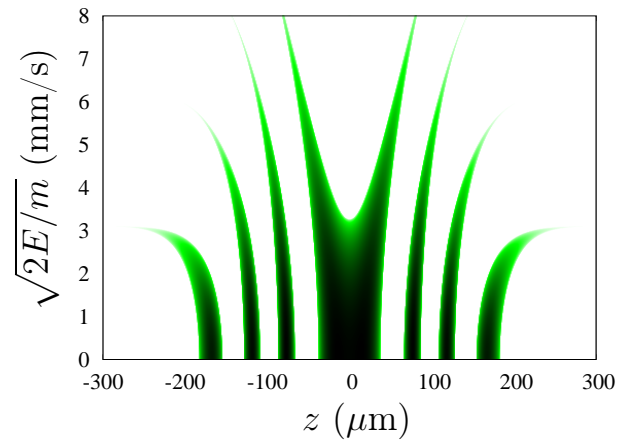}
\caption{(Color online) Position dependent imaginary part of the Mathieu exponent, $\kappa$, as a function of the velocity $(2E/m)^{1/2}$ for a lattice with a Gaussian envelope ($w=140$ $\mu$m) and for a maximum potential $U_0=10E_{\rm R}$.}
 \label{figurev10}
\end{figure}

We have represented in Fig.~\ref{transmission}(b) the position dependent imaginary part of the Mathieu exponent for a lattice having a Gaussian envelope ($w=140$ $\mu$m) and for a maximum potential depth $U_0=E_{\rm R}$. By increasing the potential depth, we increase the number of lateral cavities. This is illustrated in Fig.~\ref{figurev10}. In principle, one could implement a more involved envelope to generate a random distribution of spatial gap barriers. The physics of Anderson localization could therefore be revisited in this context \cite{andersonnote}.

\begin{figure}[h!]
\centering
\includegraphics[width=8cm]{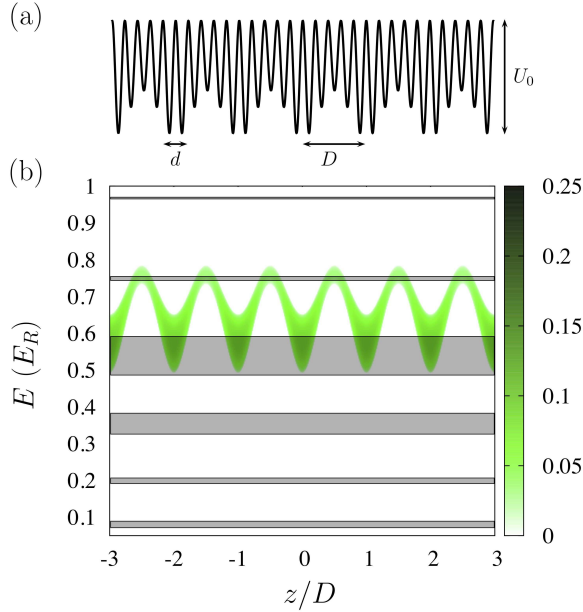}
\caption{(Color online) (a) Superlattice potential (\ref{potmod}) with $\epsilon=0.25$ and $D=5d$.  (b) The imaginary part of the wave vector obtained from the \emph{local} Mathieu equation is represented with the green/grey color code. The exact solution yields flat bands represented as black bands.}
 \label{figure5}
\end{figure}

However, one has to be careful with the modification of the picture for local spatial gaps due to multiple interferences as already discussed in Sec.~\ref{secfpcavity}. To put forward this point, we 
consider the specific case of an optical lattice with a periodic envelope (see Fig.~\ref{figure5}(a)):
\begin{equation}
V_{0}(z)=U_0\left[1+\varepsilon\cos\left(\frac{2\pi z}{D}\right)\right],\label{potmod}
\end{equation}
where $U_0>0$, $D\ge d$ is the spatial period of the envelope and $\varepsilon>0$ the amplitude of the spatial modulation (see Fig.~\ref{figure5}(a) plotted for $\varepsilon=0.25$, $D=5d$). A naive approach based on the local envelope approach yields a periodic pattern for the spatial gaps. Actually, this problem can be solved exactly by noticing that the expansion of the total potential $U(z)$ contains nothing but a superposition of four periodic potential of spatial periods  $d$, $D$, $Dd/(D\pm d)$. For the sake of simplicity, we consider that the spatial periods are commensurable. In this case, the corresponding band structure can be readily calculated using the appropriate Brillouin zone and the polychromatic Hill method \cite{whittaker_watson,McLachlan_mathieu}. In Fig.~\ref{figure5}(b), we have represented the forbidden band gaps resulting from such a calculation as black bands. They are completely flat in contrast to the intuition based on the picture proposed by the local approximation. This feature results from the fact that the envelope is periodic and therefore spatially infinite. This shows the limit of the local approach.

\subsection{Phase space engineering}
\label{pseng}

\begin{figure}[h!]
\centering
\includegraphics[width=8cm]{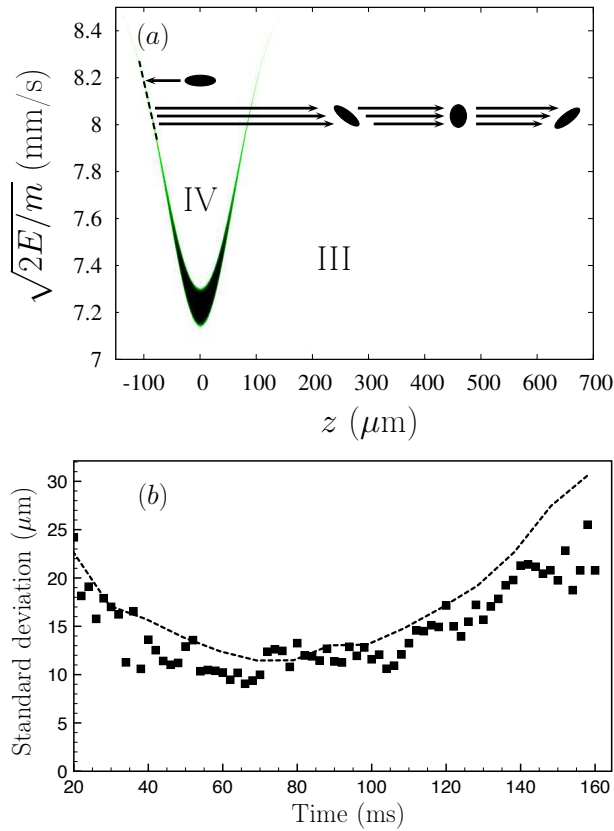}
\caption{(Color online) (a) Sketch of the emission of a wave packet through a spatial gap tunnel barrier. (b) Evolution of the standard deviation of the size of the emitted wave packet as a function of time: Experimental results (black squares), numerical result (dashed line). The numerical results take into account the finite optical resolution of the experimental imaging system (15 $\mu$m). }
 \label{figuretot}
\end{figure}

Spatial gaps enable one to generate non trivial phase space correlation. We provide a concrete example based on our experimental results. The experiment has been described in detail in \cite{dgoepl13}. In brief, a rubidium Bose-Einstein condensate is prepared in a magnetic state $F=1, m_F=-1$ in an off-resonance crossed dipole trap. The vertical beam of the dipole trap is switched off and the BEC is launched  in the horizontal optical guide provided by the other arm of the crossed dipole trap at a mean velocity $\bar{v}=8.2$ mm/s using a magnetic pulse. A finite size optical lattice having a Gaussian envelope ($w=$140 $\mu$m and depth $2.5E_{\rm R}$) and centered about the atomic wave packet is ramped on adiabatically in 1 ms just after the magnetic launching. The BEC further propagates in the lattice while being confined transversally by the horizontal guide. After 20 ms, a wave packet is emitted in the opposite direction to that set by the launching procedure. This wave packet originates from the partial tunneling through a lateral spatial gap of a wave packet that has been previously partially reflected by the other symmetric spatial gap (see Fig.~\ref{figuretot}(a)). The size of the cavity delimited by the spatial gaps increases with the energy of the wave packet. However the corresponding increase in velocity is not sufficient to overcome the increase in size and the high energy part of the wave packet is emitted after the low energy part. As a result the emitted wave packet acquires a non trivial phase space correlation (see the qualitative picture of this effect in the phase space representation of Fig.~\ref{figuretot}(a)). The experimental signature of this correlation is the focussing of the packet at a few hundreds of micrometers from the spatial gap followed by a subsequent defocussing of the packet. Figure \ref{figuretot}(b) summarizes our experimental results in which this effect can be clearly observed. A good agreement is observed with the numerical result that include the finite optical resolution of the experiment. The correlation in phase space of the emitted packet results from the variation of the size of the cavity as a function of the energy, a parameter that can be tuned experimentally by shaping the envelope of the lattice \cite{dgoepl13}. Spatial gaps therefore appear here as a way to realize a lens in time, and provide a new tool for atom optics.

\section{Spatial gap structures in higher dimensions}
\label{highdim}

Up to now, we have studied the presence of spatial gaps and their applications in one dimension. However, it is possible to generalize this concept to higher dimensions. In this section, we show how the walls provided by an appropriate shaping of the envelope of an optical lattice can be used to design matter wave cavities in higher dimensions, with new dynamical features absent in dimension one.

\begin{figure}[h!]
\centering
\includegraphics[width=6cm]{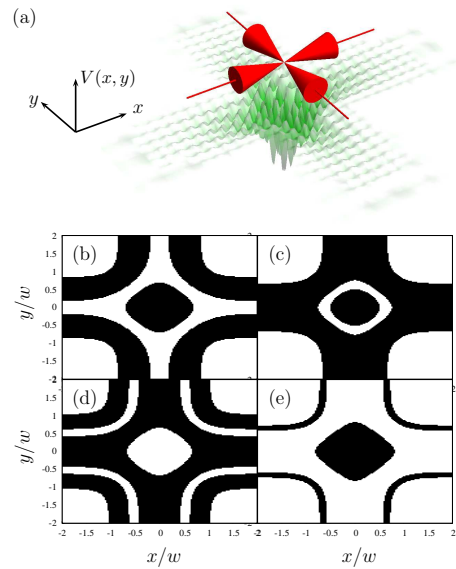}
\caption{(Color online) (a) Sketch of the potential (\ref{pot2D}) for a square lattice realized with Gaussian beams (see text). The Gaussian parameters are $x_R/w_0=y_R/w_0=20$. (b-e) Spatial gaps (black) for different values of the pseudo-energy $E_0$ and the potential depth $U_0$: (b) $E_0=0.05E_{\rm R}$ and $U_0=E_{\rm R}/2$ (c) $E_0=0.25E_{\rm R}$ and $U_0=E_{\rm R}/2$, (d) $E_0=0.35E_{\rm R}$ and $U_0=E_{\rm R}$ (e) $E_0=0.5E_{\rm R}$ and $U_0=E_{\rm R}/2$. \label{2D}}
\end{figure} 

Consider for instance a 2-dimensional square lattice potential generated by two pairs of orthogonal and counter propagating laser beams (see Fig.~\ref{2D}(a)). We assume that there is no phase relation between orthogonal beams. As a result of the Gaussian envelope of each beam, the potential experienced by the atoms reads: 
\begin{align}
&U(x,y)=-\frac{U_0}{1+\displaystyle\frac{y^2}{y_R^2}}\exp\left[-\frac{2x^2}{w_0^2\left(1+\displaystyle\frac{y^2}{y_R^2}\right)}\right]\sin^2\left(\frac{\pi y}{d}\right)\nonumber\\
&-\frac{U_0}{1+\displaystyle\frac{x^2}{x_R^2}}\exp\left[-\frac{2y^2}{w_0^2\left(1+\displaystyle\frac{x^2}{x_R^2}\right)}\right]\sin^2\left(\frac{\pi x}{d}\right).
\label{pot2D}
\end{align}

Figures~\ref{2D}(b) to (e) represent different matter wave cavities (generated by the same laser configuration) delimited by spatial gaps for different pseudo-energies and potential depths. To determine the shape of those cavities, we proceed in the following manner: (i) at each position we calculate the 2D band diagram associated with the corresponding local depth, (ii) we fix a pseudo-energy $E_0$, and (iii) we plot a black dot if the pseudo-energy lies in a gap. This representation gives a direct insight on the shape of the cavities that can be designed in 2D. Actually in 2D the Mathieu exponent becomes an anisotropic vector at each position and cannot therefore be plotted on a 2D diagram. Interestingly, cavities with different topologies can be designed. In Fig.~\ref{2D}, we give a few examples of simply (d) and non-simply (b), (c) and (e) connected cavities depending on the pseudo-energy and potential depth. 

Cavities such as the one seen in Fig.~\ref{2D}(d) could be used to probe effects of complex dynamics and tunneling. Indeed, such cavities are akin to billiards with
barriers instead of infinite walls. The shape of the cavity is known to control the dynamics, allowing the possibility to reach chaotic dynamics in parts of
the phase space, a feature which is not present in dimension one. This makes such cavities interesting to simulate the effects of chaos with cold atoms in a
closed two-dimensional system, in contrast with
the studies in \cite{BGDGO} which concerned open systems where the chaotic effects where limited by the short time spent in the chaotic region. Beyond
the study of quantum chaos in a new setting, the presence of tunable tunnel barriers at the frontier of the cavity
opens the way to new types of experimental studies of tunneling in presence of chaos (chaotic tunneling, chaos assisted tunneling) \cite{tunnelchaos}. Such effects have been
demonstrated to be important in real systems but have been difficult to observe experimentally in full detail. The versatility of the setup described here
could then allow to characterize this regime in great detail by varying the different parameters of the system.

We also emphasize that this technique provides a new method to generate curved wave guides \cite{Boshier,Phillips,Zoran} as illustrated in Fig.~\ref{2D}(d). The geometry of Fig.~\ref{2D}(b) is also well adapted for designing matter wave guided structure with two symmetric paths for interferometric purposes. Interestingly the quality of the symmetry is here automatically ensured by the beam shape. Finally, let us point out that the generalization in 3D of cavities delimited by spatial gaps would correspond in the small size regime to realizing a quantum dot for atoms, similarly to the quantum dots for electrons in mesoscopic physics.

\section{Time-dependent band gap structures}
\label{floquet}

The studies and applications presented in the preceding sections used static potentials. However, it is possible to modulate the potentials in time, leading to new effects with interesting applications \cite{Cheiney13,Arlt}.
In this section, we discuss in detail the concept of spatial gaps in an optical lattice with a periodic time-dependent amplitude:
\begin{equation}
U_{tm}(z;t)=[1+\alpha\sin^2(2\pi\nu t)]\ U(z),\label{potmod2}
\end{equation}
where $\alpha\le1$ is the strength of the modulation and $\nu$ its frequency. The solution of the Schr\"odinger equation can be expressed locally in terms of the commonly called Floquet-Bloch function \cite{Cheiney13}. Physically, the modulation opens new gaps. In the perturbative regime (small amplitude of modulation), the modulation at a frequency $\nu$ generates a new gap when $h\nu$ matches the difference $\Delta E$ of energy between two different bands of the band diagram obtained in the absence of modulation. We will consider the scattering of a non interacting wave packet onto the time-modulated potential, studying in turn the case of the wave packet initially outside and then inside the optical lattice.


\subsection{Initial wave packet outside the modulated optical lattice}

In this subsection, we thus consider a wave packet initially outside the potential (\ref{potmod2}) with $U_0=2E_{\rm R}$ and a waist $w=140$ $\mu$m. The incident wave packet has a mean velocity $\bar v=0.6 v_{\rm R}$ and a velocity dispersion $\Delta v=2.34\times10^{-4}$ m.s$^{-1}$ ($\Delta v\ll \bar v$). With such parameters and in the absence of time modulation ($\alpha=0$), the wave packet is totally reflected (the probability of transmission is lower than 1 $\%$). To further characterize the time evolution of the wave packet, it is convenient to use the semiclassical picture according to which the wave packet is described in terms of a fictitious particle having the same mean velocity \cite{Cheiney13}.

\begin{figure}[h!]
\centering
\includegraphics[width=8cm]{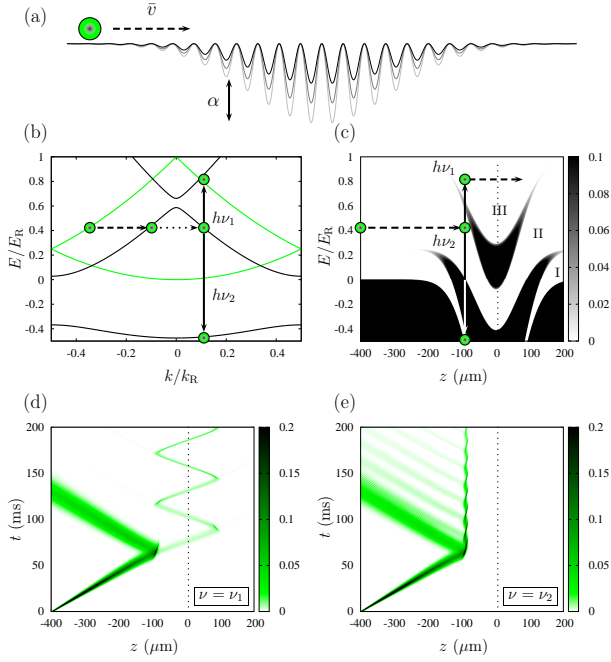}
\caption{(Color online) (a) Wave packet scattering (mean velocity $\bar v=0.65v_{\rm R}$) on an amplitude modulated lattice (\ref{potmod2}) of depth $U_0=2E_{\rm R}$. (b) Band diagram in the vanishing depth limit (green/grey line) and for a local depth $V_0(z=-104\;\mu{\rm m})=0.9E_{\rm R}$ (black line). Green/grey dots indicate the transfer between bands induced by the propagation in the presence of modulation. The wave packet mean energy of the wave packet is $m\bar v^2/2=0.42E_{\rm R}$. The solid line black arrows depict the interband resonance transitions.
(c) Position dependent imaginary part of the Mathieu exponent. The wave-packet scatters on the spatial gap (II $\rightarrow$ III). (d-e) Propagation of a non-interacting wave packet (mean velocity $\bar v$, velocity width $\Delta v=2.34\times10^{-4}$ m/s initially at $z=-400\ \mu$m) obtained from a numerical integration of the Schr\"odinger equation based on the split-step Fourier method: (d) with a modulation of frequency $\nu_1=0.3$ kHz and (e) $\nu_2=3$ kHz (modulation amplitude $\alpha=50\ \%$).\label{figmod}}
\end{figure}

Far from the potential, the mean energy of the fictitious particle is purely kinetic and equal to $m\bar v^2/2$. This sets the value of the pseudo-energy once inside the lattice. The adiabatic evolution when the fictitious particle enters the lattice is represented by the dashed black arrows (Fig.~\ref{figmod}(a), (b) and (c)). Choosing an appropriate value for the frequency modulation ($\nu_1=0.3$ kHz), one can promote the particle to the upper band in order to feed the main cavity centered about $z=0$ (see Fig.~\ref{figmod}(c)). Alternatively, one can use a larger frequency ($\nu_2=3$ kHz) to drive the fictitious particle into a lower band so as to populate a lateral cavity centered about $z=-100$ $\mu$m for our parameters (see Fig.~\ref{figmod}(c)). 
The resonance transitions between bands occur at well-defined positions as a result of the envelope and lead to the loading of the cavity only in the case of a negative velocity i.e. once the atoms have been reflected by the gap since the transition connects two energy bands with opposite slopes (the local velocity is directly proportional to the local slope of the band in the semiclassical picture \cite{Cheiney13}). 

The probability of transition increases with the modulation amplitude $\alpha$. According to our numerical simulation, for $\nu_1=0.3$ kHz and $\alpha=0.5$, 10 $\%$ of the atoms of the incoming wave packet are coupled to the main cavity delimited by the gap between bands II and III. The width of the cavity is about $200\ \mu$m for our parameters. Figure~\ref{figmod}(d) shows clearly the large oscillation amplitude of the packet inside the main cavity. This oscillation is accompanied with a focusing of the trajectories that results from the velocity dependent coupling to the cavity. With $\nu_2=3$ kHz, one transfers the atoms in the small size lateral cavity corresponding to band I (see Fig.~\ref{figmod}(b) and (c)). We observe periodic losses due to the fact that the modulation is maintained and drives regularly (at a given position) the outcoupling of atoms from the cavity. By stopping the modulation (after 75 ms), one could keep many atoms in the lateral cavity for a very long amount of time since the spatial gaps are very large and the tunneling rate is therefore completely negligible. 

\subsection{Initial wave packet inside the modulated lattice}

\begin{figure}[h!]
\centering
\includegraphics[width=8cm]{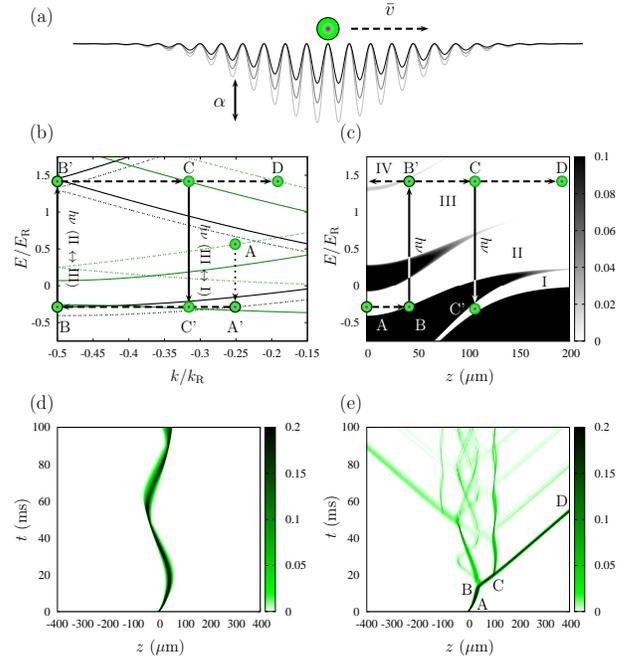}
\caption{(Color online) (a) Wave-packet propagation (mean velocity $\bar v=0.75v_{\rm R}$) inside an amplitude modulated lattice (\ref{potmod2}) of depth $U_0=2E_{\rm R}$.  (b) "Position" of the wave packet on the band diagram for different potential depths (i.e. different positions $z$). Adiabatic loading: from $U_0=0$ ($z=0$) (green/grey dashed line) to $U_0=2\ E_{\rm R}$ (black dashed line) in $t=1$ ms. Propagation: $V_0(z\simeq40\ \mu {\rm m})=U_0\simeq1.7\ E_{\rm R}$ (black line) and $V_0 (z\simeq100\ \mu{\rm m})=U_0\simeq0.9\ E_{\rm R}$ (green/grey line). The dotted arrow represents the evolution of the pseudo-energy during the adiabatic loading of the wave packet. The dashed black arrows represent the propagation of the wave packet at a constant pseudo-energy, and the solid black arrows the possible transitions between two bands induced by the modulation. (c) Position-dependent imaginary part of the Mathieu exponent. (d-e) Propagation of a non interacting wave-packet (mean velocity $\bar v$, velocity width $\Delta v=2.34\times10^{-4}$ m/s) initially at $z=0\ \mu$m: (d) in the absence of modulation and (e) in the presence of a modulation at frequency $\nu=5$ kHz, with $\alpha=33\%$. \label{figcav}}
\end{figure}

We then consider in this subsection the case of a wave packet (Fig.\ref{figcav}(a)) initially at the center of the time-dependent potential (\ref{potmod2}) and with a mean velocity $\bar v$ such that the corresponding kinetic energy is equal to the one at the middle of the second band in the vanishing potential depth limit: $m\bar v^2/2=0.562E_{\rm R}$. The potential depth is then adiabatically increased up to its final value $U_0=2E_{\rm R}$ (A$\rightarrow$A'), so that the pseudo-energy of the fictitious particle is close to -0.3 $E_{\rm R}$ (Fig.~\ref{figcav}(b)). 

In the absence of modulation, the wave packet oscillates inside the cavity delimited by the gap between bands I and II (Fig.~\ref{figcav}(d)). 
In the presence of the modulation, the wave packet propagates inside the cavity with a fixed pseudo-energy until it reaches the resonant condition for which it can be promoted to another band. For instance, with a modulation frequency $\nu=5$ kHz, the transition occurs at B. About $20\%$ of the density goes from B to B'. Actually, a more careful analysis reveals that two different paths can be followed by the atoms: (i) they are promoted on the cavity delimited by the gap that separates bands III and IV. Those atoms oscillate with the same amplitude but with a smaller time period since they have more energy and thus a larger velocity than in the original band; (ii) They are transferred to band III just below the gap (III to IV) and continue their propagation up to point C where a new resonance condition occurs. As a result of this resonance, atoms can be transferred (C to C') to the lateral cavity on the lowest band I. Part of the atoms continue their propagation (C to D) leaving the optical lattice with a mean velocity 1.2$v_{\rm R}=8.49$ mm/s, larger than $\bar v$ (these processes are illustrated in Figs.~\ref{figcav}(c) and \ref{figcav}(e)). As exemplified here, the details of the atomic wave packet evolution can be fully understood using the Bloch diagram combined with interband transitions instead of the non perturbative Floquet-Bloch diagram approach.

As illustrated above, the transitions resulting from the time modulation of the envelope offer a wide variety of possibilities to manipulate propagating wave packets. In a scattering experiment with the wave packet initially outside the lattice, the output channels can be controlled by the modulation. In the case of a wave packet placed initially at the center of the optical lattice with a finite velocity, we have shown how one can load atoms in lateral and/or centered cavities delimited by spatial gaps. The coupling between the cavities can be engineered by changing the modulation amplitude or/and the frequency. 
Furthermore a wave packet with a non-zero momentum trapped in a finite size optical lattice can be considered as an atom reservoir. Atoms can then be outcoupled at will by an appropriate choice of the  modulation frequency of the optical lattice depth. Such a device realizes a tunable source of atoms with well defined velocity properties, an important prerequisite for many quantum devices.

\section{Conclusion and perspectives} 


We have studied in detail the concept of spatial gaps created by optical lattices with a spatially varying envelope, and presented several applications for matter wave engineering and atomtronics. Indeed, we have shown that they correspond to tunnel barriers with properties difficult to attain by other techniques. In one dimension, they enable to build Fabry-Perot cavities for matter waves with tunable parameters in direct similarity with optics, to design multiple barrier systems and to shape the phase space distribution of a wave packet. In higher dimensions, we have shown that the generalization of the concept of spatial gaps makes possible to engineer curved wave guides and cavities similar to quantum dots for matter waves whose properties can be tuned. Such systems correspond to different dynamics as the parameters are varied, and enable the study of complex properties absent in one dimension such as complex dynamics and chaotic tunneling. The time modulation of the optical potential adds a new versatility to this kind of systems, enabling to load cavities or transfer atoms from one to another through the modulation.  This could allow a precise engineering of matter wave packets, in particular leading to the possibility of new tunable sources of atoms with specific velocity properties.    

In addition, spatial gaps provide a test bed for studying the role of dimensionality and interactions on tunnel effect \cite{Wu2001,Wu2003}.
In particular, the tunnel effect is deeply different in higher dimensions, enabling to probe the interplay of complex dynamics and tunneling such as
chaos-assisted tunneling or chaotic tunneling, phenomena which are absent in one dimension.  The addition of interaction effects has been the subject of very little work and could lead to new types of physical effects.  
One could also envision to revisit, in the context of a finite-size optical lattice, the dynamical instability that have been observed with a Bose-Einstein condensate in the Thomas Fermi regime, moving in a lattice \cite{Fallani04}. This instability can be explained in terms of four wave mixing \cite{Deng99} triggered by the dispersion relation inside the lattice \cite{hilligsoe05, Campbell06,Ferris08}. As a result of such processes, an atom is promoted in the high energy part of the lower band while another ends up (by energy conservation) at its bottom. Since the tunneling is favored at high energy, a cooling by coherent collisional processes could be observed in this manner.  Another perspective could be to investigate the new possibilities offered by spatial gaps using spin-dependent optical lattices \cite{DeJ98,WGF05,BFA13}.

We therefore think that the spatial gaps created by the local band gap structure of optical lattices enable to build new types of physical systems which can serve as tools for the developing field of atomtronics and can lead to the investigation of new physical effects.

\begin{acknowledgments}

We thank F. Vermersch for early discussions about this subject.  We thank CalMiP for access to its supercomputers and the University Paul Sabatier (OMASYC project). This work was supported by Programme Investissements d'Avenir under the program ANR-11-IDEX-0002-02, reference ANR-10-LABX-0037-NEXT.

\end{acknowledgments}

\begin{appendix}

\section{Perturbative treatment in the weak potential approximation}\label{perturb2}

Here, we work out a first order perturbation theory valid inside and outside the first gap which complements Sec.~\ref{section:perturbative}.
Consider the lowest gap of a periodic potential. In the weak potential depth limit, only the first two bands matters for the dynamics and the eigenvalue problem is equivalent to the two coupled equations \footnote{At the edge of the first Brillouin zone $k=\pm k_{\rm R}/2$ and we recover Eq.~(\ref{bragg}).}:
\begin{equation*}
\begin{cases}
\mathcal{E}_{k}v_{k}-(U_0/4)v_{k-k_{\rm R}}&=Ev_{k}\\
-(U_0/4)v_{k}+\mathcal{E}_{k-k_{\rm R}}v_{k-k_{\rm R}}&=Ev_{k-k_{\rm R}}
\end{cases}.
\end{equation*}
The solutions of this system are:
\begin{equation}
E^{(\pm)}_k=\frac12(\mathcal{E}_k+\mathcal{E}_{k-k_{\rm R}})\pm\sqrt{\left(\frac{\mathcal{E}_k-\mathcal{E}_{k-k_{\rm R}}}{2}\right)^2+\frac{U_0^2}{16}},
\label{A1}
\end{equation}
with $\mathcal{E}_k=\hbar^2k^2/(2m)$. Using the notation $\tilde k=k/k_{\rm R}-1/2$ and $s=U_0/E_R$, we can rewrite Eq.~(\ref{A1}) as
\begin{equation*}
\frac{E^{(\pm)}_{\tilde k}}{E_{\rm R}}=\frac{1}{4} + \tilde k^2\pm \sqrt{\tilde k^2 + \displaystyle \frac{s^2}{16}}
\end{equation*}
from which we infer the value of the wave vector as a function of the energy
\begin{equation}
k^{(\pm)}(E)/k_{\rm R}=\frac{1}{2}\left(1\pm\sqrt{1+4\frac{E}{E_{\rm R}}\pm\sqrt{16\frac{E}{E_{\rm R}}+s^2}}\right).\label{kappa}
\end{equation}
Interestingly this first order approximation formula gives the Mathieu exponent in the band gap when an imaginary part is present. 
The relative error $\epsilon$ between the exact Mathieu exponent and the first order approximation is defined by
\begin{equation}
\epsilon=\max_{0\le E \le E_R} \left( \frac{K_{\rm exact}(E)-K(E)}{K_{\rm exact}(E)} \right).
\end{equation}
The essential contribution to $\epsilon$ originates from the energies close to the gap.
We report below (Table \ref{table1}) the value of $\epsilon$ for different potential depths $U_0$. 

\begin{center}
\begin{tabular}{|c | c c c c  |}
\hline 
$U_0/E_{\rm R}$ & 0.1 & 0.3 & 0.5 & 1 \\
\hline\hline
$\epsilon$ & 0.008  & 0.042	& 0.089	 & 0.276 	 \\ \hline

\end{tabular}

{\small Table 1: Numerical values of $\epsilon$ for various potential depths $U_0$.}\label{table1}
\end{center}


\section{Perturbative treatment for the second gap}\label{perturb1}

\begin{figure}[h!]
\centering
\includegraphics[width=8cm]{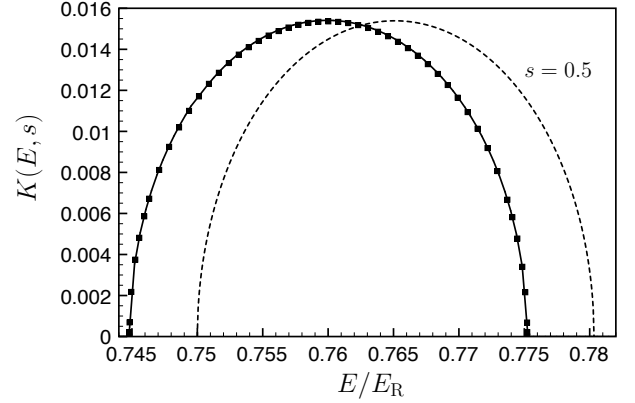}
\caption{Imaginary part of the wave vector normalized to $k_{\rm R}/2$, $K(E,s)$, in the second band gap for $s=0.5$: exact calculation (solid line), perturbative calculation to the lowest order (dashed line) and perturbative calculation pushed to the next order (black square).}
\label{fig2}
\end{figure}

Here, we extend the treatment of Sec.~\ref{section:perturbative} to the second gap.
The lower border of the second gap is located at the center of the Brillouin zone ($k=0$). By symmetry, we shall evaluate its width by considering three modes $v_0$, $v_{k_{\rm R}}$ and $-v_{k_{\rm R}}$. The energy at the edge of the gaps are obtained at the lowest order by solving the following coupled mode equations:
\begin{eqnarray} 
&& (E_{\rm R}-U_0/2)v_{-k_{\rm R}} - (U_0/4)v_{0} = Ev_{-k_{\rm R}}, \nonumber \\  
&& -(U_0/4)v_{-k_{\rm R}} - (U_0/2)v_{0} - (U_0/4)v_{k_{\rm R}} = Ev_{0}. \nonumber \\
&& -(U_0/4)v_{0} + (E_{\rm R}-U_0/2)v_{k_{\rm R}}  = Ev_{k_{\rm R}}. \nonumber \\  
\end{eqnarray}
One finds $E_-=E_{\rm R}\left( 1 - s/2\right)$ and $E_+=E_{\rm R}\left( 1-s+\sqrt{1+s^2/2} \right)/2$. In this case, the energy difference scales as $s^2$: $
(E_+-E_-)/E_{\rm R}=(\sqrt{1+s^2/2}-1)/2\simeq s^2/8$. To determine the imaginary part of the wave vector inside the second gap, we shall follow the same method as for the first gap (see Sec.~\ref{section:perturbative}). For a given energy $E_x=(E_++E_-)/2+x(E_+-E_-)/2$ with $-1<x<1$ inside the gap, we search for a solution involving the three following wave vectors $k_\pm=\pm (1\pm iK/2)k_{\rm R}$ and $q_0=ik_{\rm R}K/2$:
\begin{eqnarray} 
&& ((1-iK/2)^2-s/2-e_x)v_{-k_{\rm R}} - (s/4)v_{0} = 0, \nonumber \\  
&& -(s/4)v_{-k_{\rm R}} + (-K^2/4 - s/2-e_x)v_{0} -(s/4)v_{k_{\rm R}} = 0, \nonumber \\
&& -(s/4)v_{0} + ((1+iK/2)^2-s/2-e_x)v_{k_{\rm R}}  = 0, \nonumber \\  
\end{eqnarray}
with $e_x=E_x/E_{\rm R}$. The determinant of this three-mode system yields a polynomial in $K$ of order 6 whose unique positive real solution is the imaginary part we are interested in. In Fig.~\ref{fig2}, we represent the corresponding result for $s=0.5$. We have also plotted the result of a five-mode approximation which is in very good agreement with the exact result.

\end{appendix}


\begin{thebibliography}{99}
\bibitem{Note}
Note that the complex transmission coefficient, $te^{i\varphi_t}$, and reflection coefficient, $re^{i\varphi_r}$, for one spatial gap are energy-dependent.
\bibitem{andersonnote}
In comparison with previous works \cite{AndersonL}, the investigation of this effect using spatial gaps requires to use wave functions with a much smaller energy width than that of a Bose-Einstein condensate because of the relatively large distance between the barriers. This can be done using phase space manipulation \cite{Condon14}.
\end{thebibliography}
\end{document}